\documentclass[useAMS]{mn2e} 

\usepackage{graphicx}

\usepackage{longtable}

\title[The Age and Structure of the Galactic Bulge from Mira
Variables] {The Age and Structure of the Galactic Bulge from Mira
Variables} \author[R. M. Catchpole et al.]{Robin M. Catchpole$^1$,
Patricia A. Whitelock$^{2,3,}$, Michael W. Feast$^{3,2}$,\newauthor Shaun
M. G. Hughes$^4$, Mike Irwin$^1$ and Christophe Alard$^5$\\ $^1$
Institute of Astronomy, University of Cambridge, Madingley Road,
Cambridge, CB3 0HA, United Kingdom.\\ $^2$ South African Astronomical
Observatory, P.O.Box 9, 7935 Observatory, South Africa.\\ $^3$
Astronomy, Cosmology and Gravity Centre, Astronomy Department, 
University of Cape Town, 7701, Rondebosch, South Africa.\\ $^4$
AVEVA Solutions Ltd, High Cross, Madingley Road, Cambridge, CB3 0HB, United
Kingdom.\\ $^5$ Institut d'Astrophysique de Paris, 98bis, Boulevard
Arago, F-75014, Paris, France}
\begin{document}
\maketitle
\begin{abstract} We report periods and $JHKL$ observations for 648 oxygen-rich Mira variables found in
two outer bulge fields at $b=-7^{\rm o}$ and $l=\pm8^{\rm o}$ and
combine these with data on 8057 inner bulge Miras from the OGLE, Macho and 2MASS
surveys, which are concentrated closer to the Galactic centre. Distance
moduli are estimated for all these stars. Evidence is given showing that the bulge structure is a function of age. The longer period
Miras ($\log\,P> 2.6$, age $\sim 5$ Gyr and younger) show clear evidence of a bar structure
inclined to the line of sight in both the inner and outer regions. The
distribution of the shorter period (metal-rich globular cluster age)
Miras, appears spheroidal in the outer bulge. In the inner region these old stars are also distributed differently from the
younger ones and possibly suggest a more complex structure.
These data suggest a distance to the galactic centre $R_0$, of 8.9\,kpc with an estimated uncertainty of $\sim 0.4$\,kpc. The possible effect of helium enrichment on our conclusions is discussed.

\end{abstract}

\section{Introduction}

 Detailed studies of the infrared surface brightness of the Galactic bulge
(Matsumoto et al. 1982; Blitz \& Spergel 1991) suggested a triaxial
ellipsoid or bar with a major axis inclined to the line of sight and closer
to us at positive Galactic longitudes.  Whitelock \& Catchpole (1992) see
also (Whitelock 1992, 1993) used the distribution of individual bulge Mira
variables, with distances from a near infrared period-luminosity relation,
to show that these stars were in a bar with major axis at $\sim45^{\rm o}$ 
to the line of sight.  Work on bulge structure to about 2009 was summarized
in table 1 of Vanhollebeke et al.  (2009) (also reproduced in Rich 2013). 
This table, together with later work (e.g.  Gonzalez et al.  2011; Gerhard et al.  2012; D\'{e}k\'{a}ny et al.  2013; Pietrukowicz et al.  2012, 2015; 
Wegg et al.  2013;  Cao et al.  2013; Vasquez et al.  2013; Gran et al.  2015; 
Nataf et al. 2015) shows that a large range of bar angles have been
suggested.  These are generally in the range $\sim20^{\rm o}$ to $\sim
45^{\rm o}$ but have ranged from $11.1\pm 0.7^{\rm o}$ (Robin et al.  2003)
 to a near spherical
distribution (D\'{e}k\'{a}ny 2013).  Apart from observational uncertainties,
one reason for this large range is the way the data have been handled.  The
locus of mean distance along lines of sight to the bulge can have quite a
different slope to that of the major axis of a bar, derived from the same
data using a model (e.g.  Wegg 2013; Pietrukowicz et al.  2015).  For
instance, for RR Lyraes in a central bulge region Pietrukowicz et al.
(2015) find an angle of $56^{\rm o}$ from the mean distances in different
directions and $20^{\rm o}$ from an elliptical model.  It is also now
becoming realized that the observed bulge shape may depend on the region of
the bulge being studied as well as on the type of tracers used.

Whilst, as documented by Rich (2013), there is a large range in metallicities among
bulge stars, there has not been full agreement on the age range in the bulge. It
seems clear that there is
an old (globular
cluster age) population in the bulge from the presence of a strong
population of RR Lyrae variables and it has often been maintained that the bulge
is entirely or almost entirely
composed of a very old population (e.g. from colour-magnitude
studies of a field at
$l,b: 0.3,-6.2$ (Zoccali et al. 2003)).  Clarkson et al. (2011) put the fraction of the population
younger than 5 Gyr as at the most
3.4 percent in a field at $l,b: 2.65,-1.25$. 
However, a recent study of microlensed stars in the bulge suggests at least a
small 
population of young stars scattered through the bulge  (Bensby et
al. 2013).  
Bulge Mira variables, which show a wide range in periods and for which there is a period
age relationship, have, in fact, long suggested a significantly
younger population exists there. Nataf (2015) recently reviewed the evidence for and against the presence of a younger population and concluded that enhanced helium-enrichment would go some way to resolving the controversy.

In the present paper we report periods and $JHKL$ photometry of 648 Mira variables
in the bulge, chosen to study bulge structure. These observations are
combined with 2MASS data for 8057 bulge Miras with OGLE and/or Macho
periods, to study the structure of the bulge as a function of age as
well as to estimate the distance to the Galactic  centre.

\section{Mira Variables} \label{sec2}
Mira variables are Asymptotic Giant Branch (AGB) stars with long pulsation periods (
$P \geq 100$ days) and
large amplitude variations ($\Delta V >2.5$, $\Delta K > 0.4$ mag). They are cool
($T_{eff}<3500$K) and are generally understood to be near the end of the AGB
evolutionary phase where they are rapidly losing mass (typically $10^{-6} < \dot{M}
<10^{-4}$). Kinematic and other studies  (e.g. Feast \& Whitelock 2000; Feast 2008) have shown that the
period of a Mira is a good indication of its age and/or initial mass.
This suggests that whilst the shorter period Miras
(some of which are found in metal-rich globular clusters) are very old
with $M_i<1 M_{\odot}$, the bulk of
Miras in the solar neighbourhood with $\log P \sim 2.5$ are $\sim 7$Gyr old\footnote{logarithms are to base 10 throughout this paper}.
An age of $\sim 3$ Gyr  has been estimated at $\log P \sim 2.65$ and in the
following we refer to these as of ``intermediate age".
Longer period Miras (including OH/IR stars which are generally long period Miras)
are even younger. 
  
As shown by Wood (2000), AGB variables lie on several parallel sequences in a $K$
period-luminosity (PL) diagram,
where each sequence is a different pulsation mode. The Mira variables, which have
larger amplitudes than the semi-regular (SR) variables, fall mostly on the PL
relation corresponding to fundamental pulsation (sequence C in Wood 2000). However,
at the long periods ($P>400$ days) the mass-loss rates are sometimes sufficiently
high that circumstellar extinction forces the apparent luminosity below the PL
relation, unless the extinction is corrected for (e.g. Whitelock
et al. 1991; Glass et al. 1995; Matsunaga et al. 2009).

In the Large Magellanic Cloud we also find a few Miras above the PL relation
(e.g.  Feast et al.  1989); these are believed to be relatively high mass
hot-bottom-burning (HBB) stars (Whitelock et al.  2003, Menzies et al. 
2015) possibly pulsating in the first overtone (Feast 2009).  We expect
these higher mass objects to be concentrated toward the centre and close to
the plane.  Their numbers in our fields are expected to be small and should
not affect our discussion.

The Mira PL relation
is potentially a very useful tool for distance scale studies, within the Galaxy and
at larger distances, as recently summarized by Whitelock (2013).
 However, correction for both interstellar and circumstellar extinction must be made.
This is discussed in section \ref{ext}.
The intrinsic colours of Galactic long period variables
were discussed by Feast et al. (1982) and using these data Glass \& Feast (1982)
isolated an area of a $JHK$ two-colour diagram (their fig. 1) in which dereddened
Miras are to be found. This is the referred to as the ``Mira box" in the following
discussion.

The atmospheric abundance of carbon relative to oxygen determines whether Miras are oxygen rich ($\rm C/O<1$) or carbon rich ($\rm C/O>1$). Carbon enrichment happens during third dredge-up (Iben \& Renzini 1984) in stars which have sufficiently large initial masses, and the existence or otherwise of carbon stars has sometimes been taken as evidence for or against the existence of an intermediate mass population. Note that more carbon must be dredged-up in order to produce carbon stars in a population that has a high initial oxygen abundance, as does the bulge, and that dredge-up efficiency depends on the initial abundance as well as on the initial mass, being less efficient at high metallicity or high helium abundance (Karakas et al. 2014 and references therein).  The process of hot bottom burning (e.g. Boothroyd, Sackmann \& Ahern 1993) prevents intermediate mass stars from becoming carbon rich, particularly at low metallicity, although this is probably not relevant to the Galactic bulge.

It should be noted that we assume that the Miras in the bulge are all or nearly all
oxygen rich. Optical spectroscopy of giants suggests there are only a very few
C-rich stars in the bulge (Blanco, McCarthy \& Blanco 1984) and those
few are faint, blue and probably the products of binary
evolution. More recently Soszynski et al. (2013) have interpreted the
OGLE light-curves of AGB stars as supporting an O-rich character for
the majority of the variables.
Relatively little is known about the chemistry of highly obscured
variables towards the Galactic centre, except for those from which
OH-Masers have been detected (e.g. Sevenster 1999), which are
obviously O-rich. Cole \& Weinberg (2002) are sometimes quoted as
demonstrating the presence of numerous C-stars in the bar of the
bulge, on the basis of 2MASS colours (and some unpublished spectra of
the inner Galaxy). The interpretation of the 2MASS colours is based on
the assumption that C stars have large $J-K$ colours, as is found in
lower metallicity environments such as the LMC.  However, the $JHK$
colours of mass-losing O-rich stars are very similar to those of
C-stars and these will dominate in the higher metallicity Galactic
bulge.  Ojha et al. (2007) have questioned the Cole \& Weinberg
interpretation  of the 2MASS colours and  Ishihara et al. (2011) show,
from a study of stars with both 2MASS and Akari photometry, that the
majority of AGB stars in the direction of the Galactic centre are
O-rich. This is confirmed by Uttenthaler et al. (2015).

\begin{figure} 
\includegraphics[width=8.5cm]{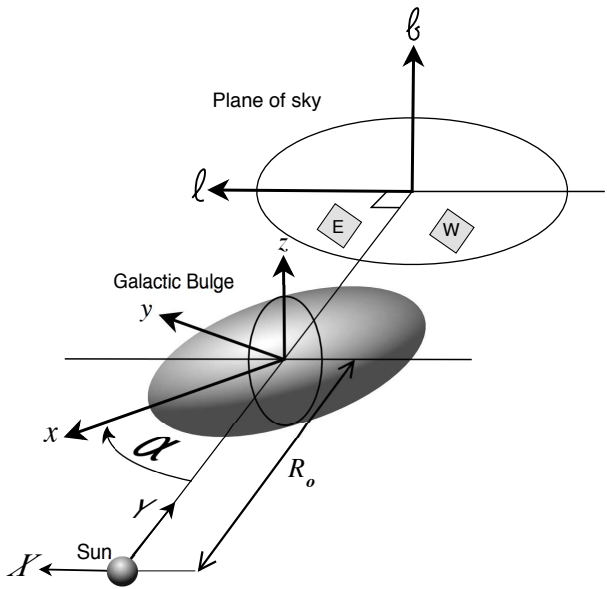}
\caption{Shown, are the relative location of our two fields and the definition
of the various coordinate systems, including the angle ($\alpha$) of
the bulge to the Sun-centre line of sight. Compare this with
Fig.~\ref{allstar} to see the distribution of the Miras. }
\label{bulgemod}
\end{figure}

\section{Previous work on bulge Miras}
The early optical survey of bulge Miras by Gaposchkin (1955), on plates taken by
Baade and that of
Oosterhoff et al. (1967, 1968), on plates taken by Thackeray and colleagues,
as well as the work of Lloyd Evans (1976),
indicated a broad range in period with most Miras having periods in the range
150 to 450 days in the NGC\,6522, Sgr\,I and Sgr\,II
 windows ($l,b;+0.9,-3.9;+1.4,-2.6:+4.2,-5.1$).
Most of the bright IRAS sources delineating the overall structure of the bulge
are Mira variables (Feast 1986; Glass 1986)
and these tend to have even longer periods (Whitelock et al. 1991; Glass et
al. 1995), with a few as long as $\sim 700$ days.  
 It is possible that some of the longest period stars may be due to binary
mergers as suggested by Renzini \& Greggio (1990) (one probable example of
such a process is a carbon rich Mira with a period of 551 days in a globular
cluster (Matsunaga 2006; Feast et al.  2013).  However, it seems very
unlikely that this could be the source of all the 300 to 400 day Miras (see
also section \ref{sec2}).  Thus, taking the presence of long period Miras
together with the discussion in section \ref{sec2}, indicates a substantial
intermediate age population in the bulge.

As mentioned above, Whitelock \& Catchpole (1992) and Whitelock (1992)
observed 141 IRAS Miras in two strips on either side of the Galactic centre
at $
-15^{\rm o}<l<15^{\rm o}$ and $7^{\rm o}<|b|<8^{\rm o}$ and made the first
attempt to use individual stars to probe the structure of the bulge in three
dimensions.  A good fit to their data was given by a triaxial bulge inclined
at $45^{\rm o}$ to the line of sight.  The relatively high Galactic latitude
of the fields indicated a thick bar and there were suggestions of an ``X"
structure.

Sevenster (1999) discussed the distribution and velocities of 507
OH/IR stars within $10^{\rm o}<l<-45^{\rm o}$ \& $|b|<3^{\rm o}$. She
finds evidence for an inclined bar as well as other structures. She
also finds 9 stars from her young sample, lying along an HI feature
identified with the 3 kpc arm.  Groenewegen \& Blommaert (2005) use
DENIS and 2MASS colours to analyse 2691 Mira variables identified in
the OGLE Survey. From these data they derive mean distances along
various lines of sight and find evidence for a bar at an angle of
$47^{\rm o}$ to the line of sight.

 Matsunaga, Fukushi \& Nakada (2005) find 1968 Mira variables in the OGLE II data
base. They assign them 2MASS colours and note that their surface
density is similar to that of the 2.2$\mu$m COBE image. The radial
distribution shows a tilt towards us in positive longitudes.  The work of
Matsunaga et al. (2009) on Miras close to the centre is discussed in section \ref{distance}.

\section{Observations} 

We selected two fields (the SAAO fields), both at $b=-7^{\rm o}$, centred at
$l=+8^{\rm o}$ (bulge East) and $l=-8^{\rm o}$ (bulge West) and each
covering about 25 square degrees (see Figs.~\ref{bulgemod} \&
\ref{allstar}).  Over a period of about six years, the AAO Schmidt was
used to obtain IVN ($I$-band) plates in each field, which are listed in
Table~\ref{tab_dates}.  The Schmidt plates were scanned, using both the APM
in Cambridge and the MAMA at the Paris Observatory, to find the long period
variable candidates and their periods. 
 The APM periods were found using phase dispersion minimization
 (Stellingwerf 1978) and the MAMA periods using the the method described by Schwarzenberg-Czerny (1996).
 The candidates were then observed
with the SAAO 1.9-m telescope to obtain $JHKL$ photometry in the SAAO system
(Carter 1990). The error of individual observations is estimated to be
0.03 in $JHK$ and 0.05 in $L$.

\begin{table}
\caption{Schmidt Survey Plates}
\label{tab_dates}
\begin{tabular}{cccc} \hline \multicolumn{2}{c}{East} &
\multicolumn{2}{c}{West}\\  \multicolumn{2}{c}{Field 1834 -2500}&
\multicolumn{2}{c}{Field 1754 -4042}\\  \hline
\multicolumn{1}{c}{Plate No.} &
\multicolumn{1}{c}{Date.}&\multicolumn{1} {c}{Plate No.} &
\multicolumn{1}{c}{Date}\\ \hline I14563 & 27/09/91 & I14565 &
28/09/91\\ I14624 & 26/10/91 & I14609 & 14/10/91\\ I14862 & 29/03/92 &
I14631 & 28/10/91\\ I14975 & 22/05/92 & I14852 & 27/03/92\\ I15402 &
18/03/93 & I14903 & 22/04/92\\ I15553 & 15/05/93 & I14909 & 24/04/92\\
I15652 & 28/07/93 & I14974 & 22/05/92\\ I16059 & 07/05/94 & I15039 &
02/07/92\\ I16147 & 15/06/94 & I15093 & 29/07/92\\ I16162 & 28/06/94 &
I15134 & 25/08/92\\ I17032 & 14/04/96 & I15151 & 06/09/92\\ I17081 &
09/05/96 & I15377 & 13/03/93\\ I17155 & 05/08/96 & I15481 & 16/04/93\\
I17644 & 13/07/97 & I15552 & 15/05/93\\ I17716 & 08/09/97 & I15592 &
14/06/93\\ &  & I15630 & 17/07/93\\ &   & I16051 & 04/05/94\\ &  &
I16071 & 10/05/94\\ &  & I16134 & 03/06/94\\ &  & I17026 & 28/03/96\\
\hline
\end{tabular}
\end{table}

\begin{table*}
\caption{$JHKL$ photometry on the SAAO system for Mira
variables found in the SAAO fields. The errors of observation are 0.03
magnitudes in $JHK$ and 0.05 in $L$. The values of $A_v$ are obtained by
the Whitelock method. See text for details. The full table is
available online.}
\label{mira}
\begin{tabular}{cccrcrrrrl} \hline
\multicolumn{1}{c}{RA} & \multicolumn{1}{c}{Dec} &
\multicolumn{1}{c}{l} &\multicolumn{1}{c}{b} & \multicolumn{1}{c}{P}
&\multicolumn{1}{c}{$J$} & $H$ & $K$ & $L$ &\multicolumn{1}{l}{$A_v$}
\\ \multicolumn{2}{c}{(2MASS equinox 2000)}& \multicolumn{2}{c}{degree} &
\multicolumn{1}{c}{day}& \multicolumn{4}{c}{magnitude}\\ \hline
 265.439978& -43.497269& 347.14&  -6.93& 245.1&  9.94&  8.89&  8.44&  7.96&  1.17 \\ 
 265.443089& -41.124081& 349.18&  -5.69& 227.1& 10.30&  9.25&  8.80&  8.34&  1.40 \\ 
 265.512191& -41.170151& 349.17&  -5.76& 239.1&  9.81&  8.82&  8.39&  7.68&  1.40 \\ 
 265.513947& -39.581005& 350.53&  -4.94& 257.1&  9.41&  8.40&  7.92&  7.19&  1.89 \\ 
 265.542028& -39.615677& 350.51&  -4.97& 353.1&  9.38&  8.26&  7.62&  6.97&  1.82 \\ 
 265.543751& -41.269283& 349.09&  -5.84& 271.1&  9.29&  8.35&  7.86&  7.19&  1.44 \\ 
 265.548023& -38.449688& 351.51&  -4.37& 427.2&  9.59&  8.05&  7.23&  6.37&  3.20 \\ 
 265.555889& -39.809402& 350.35&  -5.08& 347.1&  8.56&  7.45&  6.84&  6.20&  1.79 \\ 
\hline
\end{tabular}
\end{table*}

\begin{figure} 
\includegraphics[width=8.4cm,angle=0]{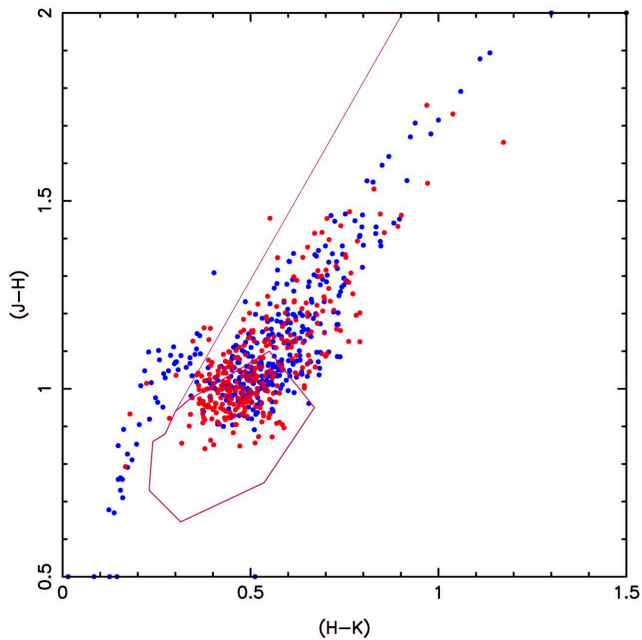}
 \caption{Two-colour plot for all our candidate stars. The Miras are
 defined as those stars which lie inside the box or below the straight line
   defining its reddened extension. Miras in bulge East, with positive
   longitudes, are shown in blue and those in bulge West, in red. }
 \label{jhhk}
\end{figure}

\begin{figure} 
\includegraphics[width=8.6cm,angle=0]{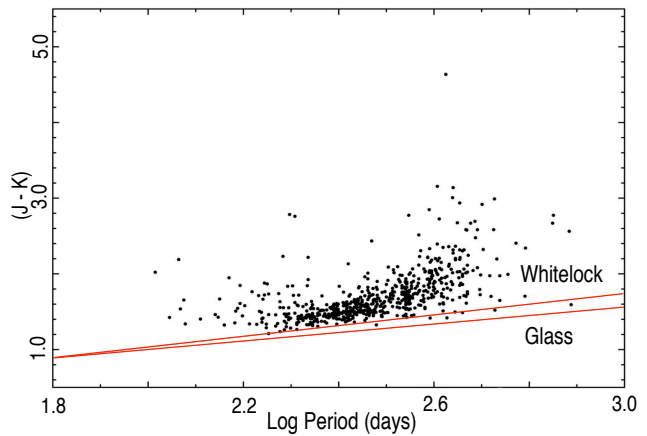}
 \caption{Period-colour relation for our Miras listed in
   Table~\ref{mira}. The line labelled Whitelock is the locus for
local Miras given in Whitelock et al. (2000). The other is for LMC Miras 
from Glass et al. (1995).}
 \label{jklogp}
\end{figure}

 A $(J - H)$, $(H - K)$ two-colour diagram, Fig.~\ref{jhhk}, was then used
to distinguish between Miras and non-Miras. 
The region for unreddened local Miras
is defined in this diagram by the Mira box (Feast et al. 1982)
 and all stars lying
in this region and to the right of the line defining its upper reddened extension, are listed as Miras in
Table~\ref{mira}.
The Miras are shown in a $(J-K)$/$\log P$ diagram in
Fig.~\ref{jklogp}. The two straight lines are the locus of unreddened
Miras discussed below.

 Although most stars have only two observations 28 were observed an
average of 10 times each to confirm the periods found from the
photographic plates. 

The magnitudes listed in Table~\ref{mira}  are mean values.
All the stars have been identified with stars in the 2MASS catalogue
and assigned 2MASS positions. These are listed in
Table~\ref{mira}. 

Ten stars in bulge West and 51 in bulge
East, lie above the reddening line in Fig.~\ref{jhhk}. Dereddening will move them
into the region of normal giant stars and they were omitted from the analysis.

The greater
proportion of non-Miras in the East field probably arises from 
identification problems when carrying out observations at the telescope.
This leaves a total of 341 Miras in bulge West and 307 in bulge East.

The OGLE survey (Soszynski et al. 2013) found 6528 Mira variables in the Galactic bulge region. A total of 1093 Mira variabes were also
found in this region in the MACHO survey and are listed by Bernhard \&
H\"{u}mmerich (2013). Of these, 142 are common to the OGLE survey as are
16 stars in the SAAO fields. Five stars are common to the SAAO survey and the
OGLE survey. Duplicates were eliminated to give a grand total of
8057 stars. The spatial distribution of these stars in the plane of the sky,
is shown in Fig.~\ref{allstar}.

Of the 21 stars common to our survey, 19 have MACHO or OGLE periods that agree
within 2 days of our periods. Star number 277.00016 in
Table~\ref{mira}, designated by its 2MASS R.A. has the largest
difference. We find a period of 450 days and the MACHO survey finds a period of 463 days.

\begin{figure}
\includegraphics[width=4.2cm,angle=270]{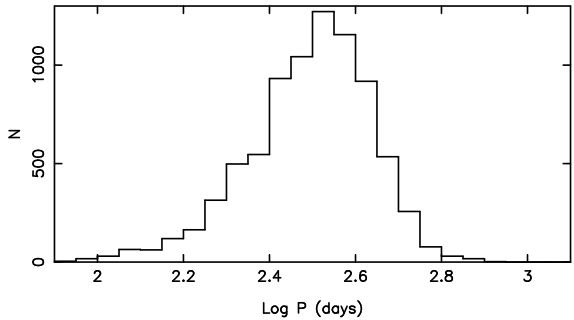} 
 \caption{A histogram showing the period distribution of the 8057 Miras discussed in this paper.}
 \label{hist}
\end{figure}

Fig.~\ref{hist} 
 shows a histogram of the periods of the bulge Miras discussed here. It shows a very broad distribution with most periods in the range $2.25 <\log P < 2.75$ ($180<P<560$ days), but with tails to the short and long period ends. We might anticipate that this will be incomplete; at the very long period end extremely red Miras (e.g. extreme OH/IR sources) could have been missed by OGLE and by us (see also Whitelock et al. 1991), while at the short period end the Miras are faint.  In globular clusters Miras have been found with periods of $ 180<P<250$ days. There are also Miras with periods around 300 days whose cluster membership requires confirmation. Miras in the solar neighbourhood cover a large range of period, but no one has yet produced a histogram for a given volume.

All the stars were identified in the 2MASS catalogue and their $J,H,K$
magnitudes transformed to the SAAO system, using the revised Carpenter
(2001) transformation formulae given on
`www.astro.caltech.edu/$\sim$jmc/2mass/v3/transformations/'. For Miras
with SAAO observations, the transformed 2MASS $J,H,K_{s}$ photometry was combined with
our photometry, raising the average number of observations from two to
three. There are uncertainties in these transformations when applied to
Miras. However, as we discuss in section \ref{distance}, analyses in the
2MASS system, are in close agreement with those in the SAAO system,
suggesting that any uncertainty in the transformations has not affected our
main conclusions.

\section{Derivation of Extinction and Distance Modulus} \label{ext}
The bulge Mira observations must be corrected for interstellar extinction
and also, at least for the longer period stars,  circumstellar
extinction. Estimates of the total reddening (interstellar and circumstellar)
were made in the following four (somewhat similar) ways:
\begin{enumerate}
\item{\bf Whitelock.} Uses the intrinsic relationship
\begin{equation} 
(J-K)_0 =-0.39 +0.71\log P 
\end{equation}
given by Whitelock et al. (2000) for local Galactic Miras. 
Note that at the longer periods this is an extrapolation (see the discussion on this point in Catchpole (1992)) and that some scatter will be introduced from the fact that we use observations from a single epoch, or small number of epochs, rather than mean colours.
However, since Miras are redder when faint, there will be 
some compensation in the final distance modulus, after correcting for extinction.

\item{\bf Glass.}  As for ``Whitelock" but using the Glass et al. (1995)
  relationship for LMC Miras.
\item{\bf JHK.} In the $(J - H)$, $(H - K)$ diagram  (Fig.~\ref{jhhk}) moving all stars parallel
  to the direction of the reddening vector until they intercept a line
  orthogonal to the reddening vector, passing through $(J -
  H)=0.87$, $(H - K)= 0.45$, which we define as the centre of the
  Mira box. This is close to correcting all the stars to
  $(H - K)_0= 1.32$. When compared with the Whitelock method, it under
  corrects the short period stars and over corrects the long period stars.

\item{\bf Matsunaga.} Derives reddenings and distance
by comparing observations with PL relations in $J,H,K_{s}$ from the
LMC (Matsunaga et al. 2009).
In using these relations we first convert the SAAO photometry to the 2MASS
system (see section 4) and adopt an LMC distance modulus of 18.49 (see below).

\end{enumerate}

In addition estimates of the interstellar reddening alone, were made
using the extinction map described by Schlegel et al. (1998) and
recalibrated by Schlafly \& Finkbeiner (2011), hereafter referred to as
Schlafly. This map describes the total
extinction through the Galaxy and should therefore, in principle be an upper limit.\\

Drimmel et
al. (2003) made a three dimensional model of Galactic extinction,
which Whitelock, Feast \& van Leeuwen (2008) used
iteratively, to correct for extinction. This model could not be used with the bulge Miras,
first,  because not all the lines of sight are included in the model and
secondly, because  many Miras lie beyond the distance limits of the
model. However, in regions of overlap the Drimmel et al. model gives
$K$ extinctions on average 0.06 magnitude greater than the Schlafly method.

Fig.~\ref{JKLOGPschle} shows the $(J-K)_0$/$\log P$ relations of Whitelock and Glass together with the bulge Miras, corrected only for interstellar reddening using the
Schlafly extinction map. The effects
of circumstellar reddening are clearly seen at the longer periods.

Extinctions in $J,H$ and $K$ were derived from the reddenings, using the reddening law given by Glass (1999).
This is close to that derived by Nishiyama et al. (2006, 2009) for stars in the direction of
the Galactic bulge.

A mean line drawn through the reddened Miras in 
Fig.~\ref{jhhk} is not exactly parallel
to the reddening line shown. This could be due to the circumstellar and/or interstellar
reddening law being different from the one adopted. On the other hand, since the dereddened
Miras are supposed to lie in the Mira box, it is quite possible that ones with very thick shells
come from positions in the box to the right of those with thinner shells. That this is the likely
explanation is shown by fig.~1 of Feast et al. (1982). The
  distribution of stars with longitude, indicated by colour in
  Fig.~\ref{jhhk}, suggest there is no systematic bias with longitude
  that will affect measurements of $\alpha$ (see below).

There may even be some contribution to the $K$ flux by emission
from their circumstellar shells, which is not accounted for, as was found by Whitelock (1987)
in symbiotic Miras (her fig.3). 
 
We note that the deviation of the Schlafly corrected Miras
from the Whitelock line in Fig.~\ref{JKLOGPschle} starts at $\log P \sim 2.5$ as
opposed to  $\log P \sim 2.7$ for the local Miras. At least some of this difference
may be due to different selection effects for the different samples.

After correction for extinction, the distance modulus of each star is
found using the relationship given by Whitelock et al. (2008)
\begin{equation} 
M_K=-3.51[\log P-2.38]-7.25 
\end{equation}
  The slope is
derived from observations of LMC Miras and the zero point was
derived from Hipparcos parallaxes of nearby Miras, VLBI
parallaxes of OH-Miras and Hipparcos parallaxes of sub-dwarfs, which
provide a distance calibration for globular cluster Miras. This calibration leads to a
distance modulus of 18.49 for the LMC.

\begin{figure}
\includegraphics[width=8.5cm, angle=0]{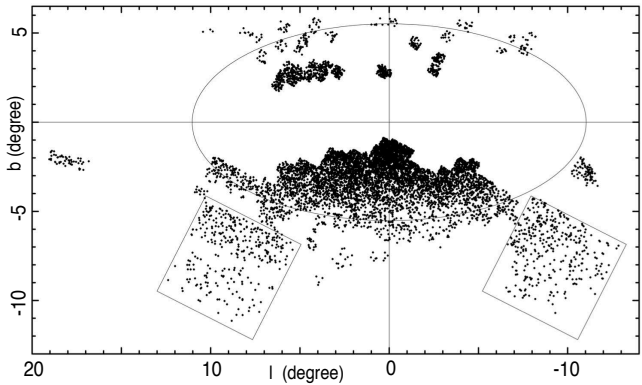}
\caption{The distribution of all the Miras in Galactic latitude and
  longitude. Miras reported here are within the two rectangles. The
  ellipse with $ a/b = 2.0$ has $ b = 800pc $ for $R_0 = 8.28\,kpc$. }
\label{allstar}
\end{figure}

\begin{figure} 
\includegraphics[width=8.6cm,angle=0]{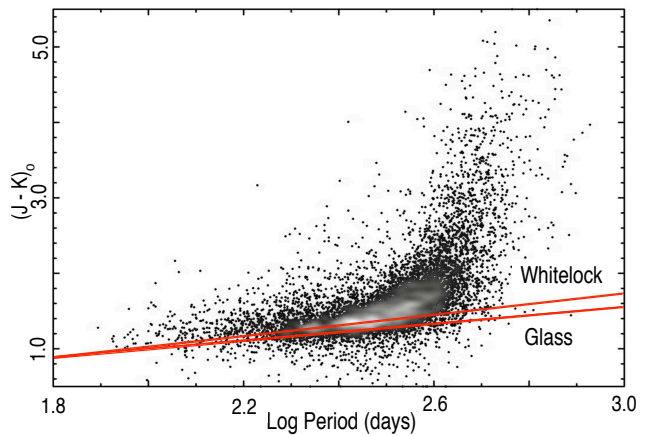}
 \caption{Period-colour relation for all the Miras, corrected for
   interstellar
   reddening, using Schlafly, as described in the text. The lines are the intrinsic lines for local
   Miras (Whitelock) and  for LMC Miras (Glass). Where the points
     are crowded, their density is represented by  grey scale contours
     with the highest density shown lightest. Note the close fit in
     this region to
     the Whitelock line up to $ \log P \sim 2.5$, where circumstellar reddening starts to dominate.}
 \label{JKLOGPschle}
\end{figure}

\section{BULGE STRUCTURE AND AGE}

\begin{table*}
\caption{Values of $R_0$ and $\alpha$, derived from a least squares fit to
  distance modulus as a function of Galactic longitude, for various intervals of
  $\log P$, Galactic latitude and longitude, using the Whitelock
  extinction, as shown in
  Fig.~\ref{ModLong} and subsequent figures. }
\label{TabRthet}
\begin{tabular}{rcccccc} \hline 
\multicolumn{1}{c}{Log Period Range}&\multicolumn{3}{c}{All
  Longitudes}&\multicolumn{3}{c}{$-3^0 \leq Longitude \le +3^o$}\\
 \multicolumn{1}{c}{day}&\multicolumn{1}{c}{$R_0$ kpc}&\multicolumn{1}{c}{\large{$\alpha$}}&
\multicolumn{1}{c}{N}
&\multicolumn{1}{c}{$R_0$ kpc}&\multicolumn{1}{c}{\large{$\alpha$}}
&\multicolumn{1}{c}{N}\\ \hline
\multicolumn{7}{c}{ $-4.5 < Latitude < +4.5$}\\
\multicolumn{1}{c}{$2.1<\log P\leq 2.2$} & $9.09 \pm 0.22$ & $118^o
\pm 17^o$ &$142$ &$8.83 \pm 0.21$ & $99^o\pm 58^o$&$87$\\
\multicolumn{1}{c}{$2.2<\log P\leq 2.3$} & $8.98 \pm 0.14$ & $94^o
\pm 14^o$ &$384$ &$8.72 \pm 0.14$ & $117^o\pm 26^o$&$255$\\
\multicolumn{1}{c}{$2.1<\log P\leq 2.3$} & $9.00 \pm 0.12$ & $101^o
\pm 12^o$ &$526$ &$8.75 \pm 0.12$ & $113^o\pm 25^o$&$342$\\
\multicolumn{1}{c}{$2.3<\log P\leq 2.4$} & $8.95 \pm 0.08$ & $86^o
\pm 8^o$ &$816$ &$8.76 \pm 0.09$ & $79^o\pm 21^o$&$494$\\
\multicolumn{1}{c}{$2.4<\log P\leq 2.5$} & $9.19 \pm 0.07$ & $85^o
\pm 5^o$ &$1579$ &$8.90 \pm 0.07$ & $78^o\pm 16^o$&$917$\\
\multicolumn{1}{c}{$2.5<\log P\leq 2.6$} & $8.83 \pm 0.04$ & $77^o
\pm 5^o$ &$1962$ &$8.72 \pm 0.07$ & $89^o\pm 16^o$&$1142$\\ \\
\multicolumn{1}{c}{$2.1<\log P\leq 2.6$} & $8.99 \pm 0.04$ & $84^o
\pm 3^o$ &$4883$ &$8.79 \pm 0.04$ & $87^o\pm 10^o$&$2895$\\
\multicolumn{1}{c}{$2.6<\log P\leq 2.7$} & $8.33 \pm 0.08$ & $49^o
\pm 4^o$ &$1182$ &$8.27 \pm 0.10$ & $61^o\pm 19^o$&$672$\\ \\
\multicolumn{7}{c}{$\mid Latitude \mid > 4.5^o$ } \\
\multicolumn{1}{c}{$2.1<\log P\leq 2.6$} & $8.78 \pm 0.08$ & $79^o
\pm 4^o$ &$1214$ &$8.97 \pm 0.12$ & $45^o\pm 14^o$&$314$\\
\multicolumn{1}{c}{$2.6<\log P\leq 2.7$} & $7.89 \pm 0.17$ & $54^o
\pm 8^o$ &$285$ &$8.54 \pm 0.24$ & $28^o\pm 13^o$&$92$\\ \\
\multicolumn{7}{c}{$\mid Latitude \mid > 6.0^o$ } \\
\multicolumn{1}{c}{$2.1<\log P\leq 2.6$} & $8.27 \pm 0.14$ & $84^o
\pm 7^o$ &$482$ &$8.36 \pm 0.27$ & $56^o\pm 42^o$&$46$\\
\multicolumn{1}{c}{$2.6<\log P\leq 2.7$} & $7.22 \pm 0.33$ & $55^o
\pm 13^o$ &$95$ &$8.71 \pm 0.40$ & $12^o\pm 4^o$&$19$\\ \\
\multicolumn{7}{c}{SAAO Fields Only } \\
\multicolumn{1}{c}{$2.1<\log P\leq 2.6$} & $8.26 \pm 0.14$ & $82^o
\pm 6^o$ &$525$ & & & \\
\multicolumn{1}{c}{$2.6<\log P\leq 2.7$} & $7.14 \pm 0.94$ & $62^o
\pm 14^o$ &$94$ & & & \\ 
 \hline

\end{tabular}
\end{table*}

 Previous work on the barred structure of the bulge has been summarized in section \ref{sec2}.
As is clear from Fig.~\ref{allstar}, the Miras in our sample at low Galactic latitude are primarily
from OGLE and concentrated toward the centre. At higher latitude the sample is dominated
by the new Miras reported in this paper, which are concentrated in
two regions in the outer bulge with mean longitudes at
$\pm 8^{\rm o}$. It is therefore convenient to discuss the low and higher latitude
variables separately.

In the following subsections we adopt the simplest model, which is to fit straight lines to various samples
of stars in distance modulus/longitude, giving unit weight to each star.  The
derived distance modulus of the Galactic centre and the angle of the distribution to the line of sight, are then converted into
$R_0$ and $\alpha$ in the XY plane.  The relation between the two systems
of coordinates is of course non-linear, but this is not of significance
for our present purpose.  Given the discussion in section 1, the angles
derived for a bar may not correspond to those derived in other ways
(particularly those derived from model fitting).  However, we are primarily
concerned with establishing whether or not various sub-samples of Miras show
evidence (or not) of barred structure and also whether there are differences
between different period groups of Mira variables.

\subsection{$\mid b \mid < 4.5^{\rm o}$.}
The sample with $\mid b \mid < 4.5^{\rm o}$ consists essentially of the OGLE stars and is concentrated
to low latitudes and longitudes in Fig.~\ref{allstar}. The stars were
divided into period groups of $\Delta \log P = 0.1$.

\begin{figure} 
\includegraphics[width=8.52cm,angle=0]{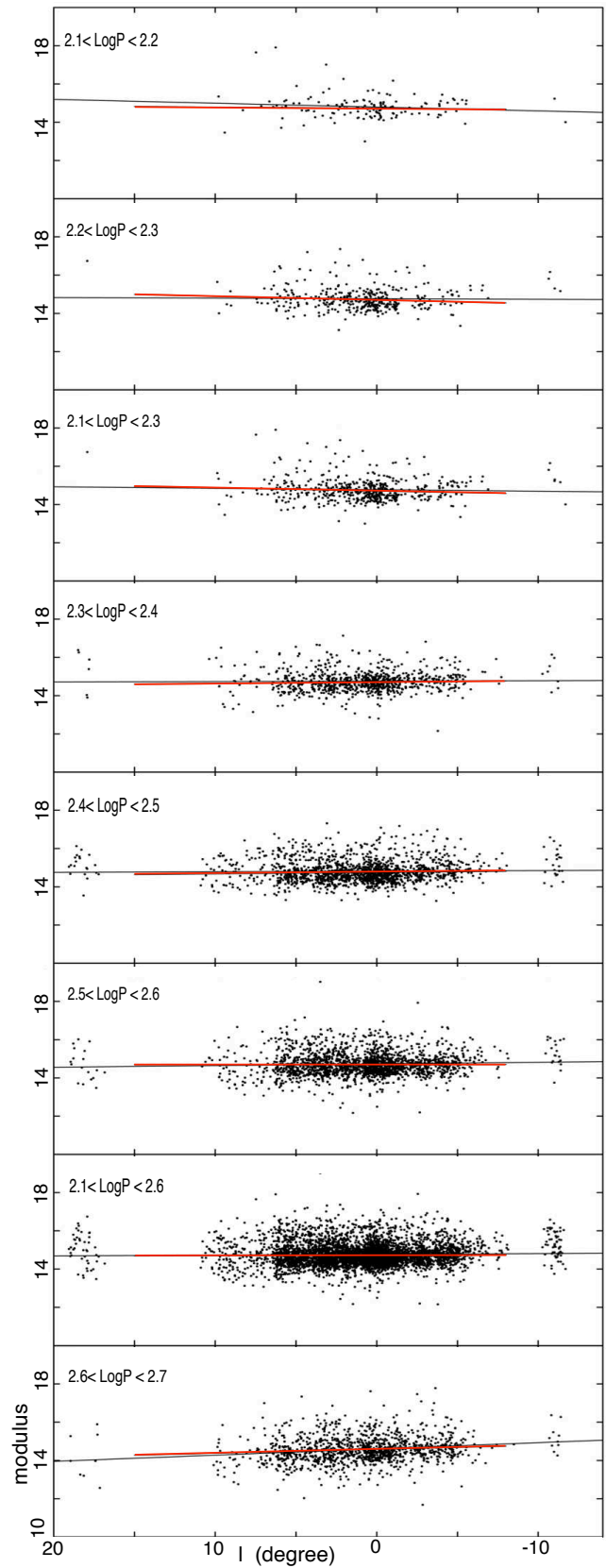}
 \caption{The variation of distance modulus with longitude for various
   intervals of $\log P$ for Miras with $\mid b \mid < 4.5^{\rm o}$. Two lines are shown. The long line is a least squares fit to
   all longitudes and the short line (red) is for $\mid l\mid < 3^{\rm o}$. The solutions are given in Table~\ref{TabRthet}.}
 \label{ModLong}
\end{figure}

\begin{figure}
\includegraphics[width=8.0cm, angle=0.0]{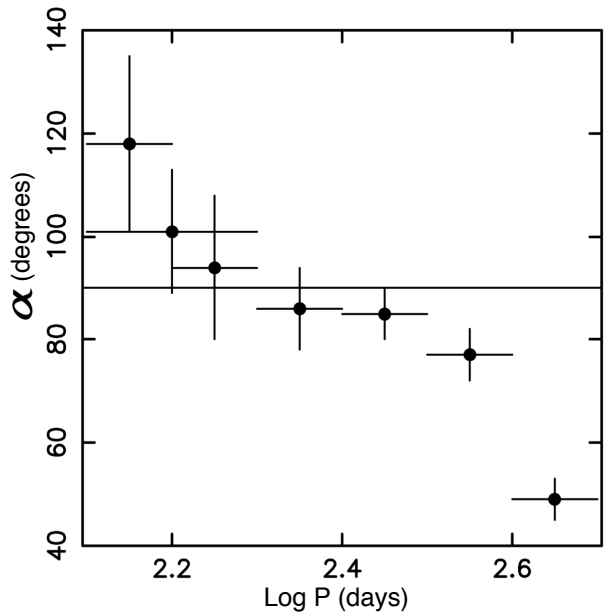}
\caption{The variation of $\alpha$ with $\log P$ for the first section of
  Table~\ref{TabRthet}. The horizontal bars give the range in log P
  and the vertical bars the standard error of $\alpha$, for each point. The Miras
  lie within $-4.5 < latitude < +4.5$ and the entire range of longitude.}
\label{thtlogp}
\end{figure}

\begin{figure} 
\includegraphics[width=8.0cm, angle=0.0]{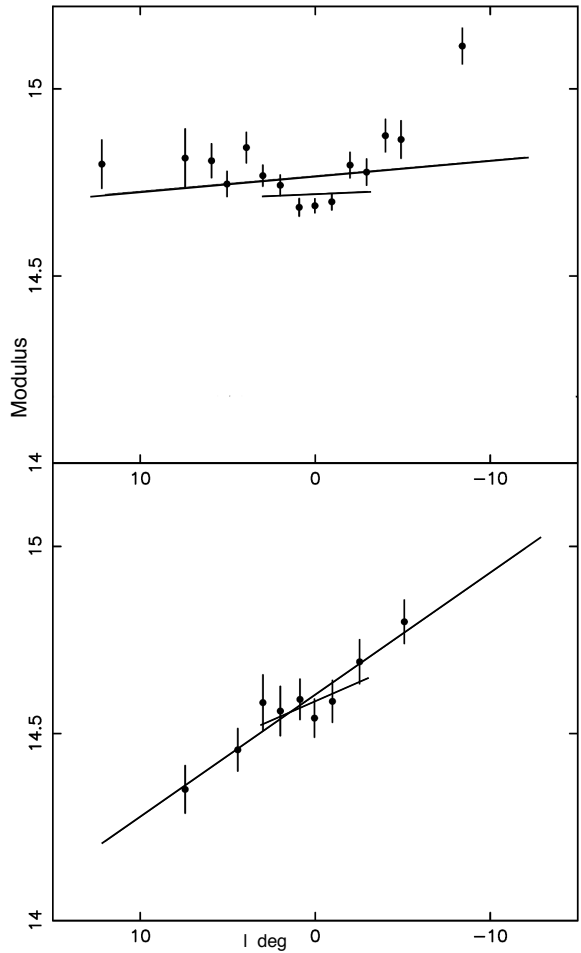}
\caption{The variation of distance modulus with Longitude for Miras with $\mid b \mid
< 4.5^{\rm
     o}$. Top, for
  $2.1< \log P < 2.6$ and bottom, for $
  2.6< \log P < 2.7$.  The least square solutions from
  Table~\ref{TabRT_Red} are shown by the straight lines. Each point
  either represents 100 Miras or a $ 1^{\rm o}$ wide longitude bin.}
\label{ModLong1}
\end{figure}

\begin{figure}
\includegraphics[width=9.0cm, angle=0.0]{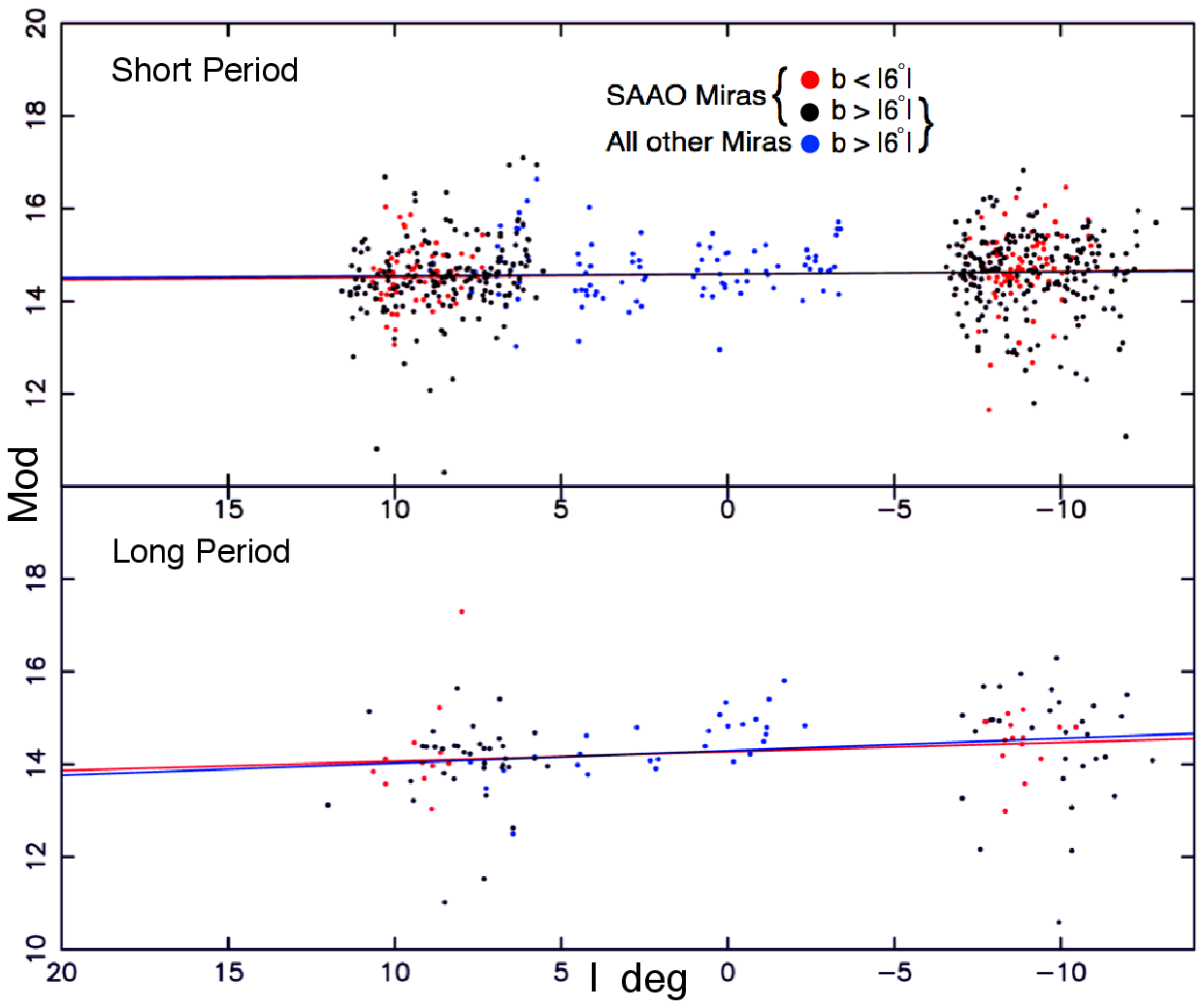}
\caption{The variation of distance modulus with Longitude, for long
  period ($2.6< \log P < 2.7$) and short
  period ($2.1< \log P < 2.6$) Miras, including all the SAAO
  Miras (all $b$) and all other Miras with $|b| >
  6^{\rm o}$ (blue). The SAAO Miras with $|b| < 6^{\rm o}$ are shown in red,
  the remainder in black. The lines are the least square solutions from Table~\ref{TabRT_Red}. The red line is the fit to the SAAO
  Miras. The blue line is the fit to all the Miras with $|b| > 6^{\rm
    o}$. For the short period Miras the lines are barely distinguishable. }
\label{p45Aug15}
\end{figure}

The results are shown in Fig.~\ref{ModLong}. The work of Nishiyama et al. (2005) (their fig. 4)
on bulge red clump stars at $b=+1^{\rm o}$ and of Pietrukowicz et al. (2012, fig. 14) on bulge RR Lyrae
variables at $-5.0^{\rm o} < b< -2.5^{\rm o}$, 
shows that the maximum number density as a function of longitude
changes slope for $\mid l \mid \sim > 3^{\rm o}$
(although in opposite directions in these two cases). We show therefore least 
square fits in Fig.~\ref{ModLong}
for all longitudes and for $\mid l\mid < 3^{\rm o}$. The values of $\alpha$ and $R_0$ derived are shown in
Table~\ref{TabRthet}. The values of $R_0$ will be discussed later.
However, it should be noted that
the $R_0$ value for the longer period stars is sensitive to the correction for circumstellar 
extinction. Note also that the values of $\sigma_{R_0}$ are internal
only.\\

Despite considerable uncertainties, Table~\ref{TabRthet} suggests  a change
in $\alpha$ with period, that is with age, which is illustrated
  in Fig.~\ref{thtlogp} for the first section of Table~\ref{TabRthet}.  This is particularly noticeable
in comparing the groups with $\log P <2.6$ with the longest period group. 
We therefore also show in Table~\ref{TabRthet} and in Fig.~\ref{ModLong}, the
results for $\log P$ between 2.1 and 2.6, which then may be compared with the
longer period group (last two rows of the first section in
Table~\ref{TabRthet}).  We deduce that the change of $\alpha$ with period
becomes marked at $\log P \sim 2.6$ (short periods, $\alpha = 84^{\rm o}\pm
3^{\rm o}$; long period $\alpha = 49^{\rm o} \pm 4^{\rm o}$).  Evidently the
long periods show clear evidence of a barred structure.  Whether the shorter
period group is barred is less clear.

According to the discussion in Feast (2008) $\log P =2.6$
corresponds to an age of $\sim 5$ Gyr.  Note that since we anticipate some
spread in the period-age relation, we cannot decide clearly whether the
results indicate a gradual change of structure with age, or the division into
two different populations at $\log P \sim 2.6$.     
A gradual change in apparent bar angle might be due to varying bar thickness
(see for instance L\'{o}pez-Corredoira et al. 2007, Appendix), it is unlikely to be a real change.       
Nataf \& Gould (2012) have suggested that the bulge contains a populations with enhanced helium abundance and Buell (2013) has discussed how this will change  evolution on the AGB. Buell suggests that the youngest main sequence stars in the bulge have enhanced helium and ages between 2 and 4 Gyr, i.e. still much younger than the Globular Clusters.  If our long period group are helium enhanced then we may not be able to deduce their ages from their periods by analogy with Miras of more normal composition.  A full discussion of this must await detailed modeling and is beyond the scope of this paper, but our conclusion that the long period Miras are younger and are in a bar seems secure.

In the rest of the paper, the terms long and short period groups, refer
to a division at $\log P =2.6$.
Fig.~\ref{ModLong1} shows the results for the short and long period groups,
but with mean points for adjacent longitude bins that either contain 100
Miras or are $1^{\rm o}$ wide.  The lines in these plots are the solutions
in Table~\ref{TabRthet}.  This allows the bulge structure to be discussed in
more detail.  The long period group clearly shows the inclined distribution
indicative of a barred structure.  However, the plot suggests a flatter
structure (larger $\alpha$) in the central region.  In view of the error bars, too much weight
should not be placed on this.  Nevertheless, this structure is reminiscent
of the distribution of red clump stars close to the Galactic equator
(Nishiyama et al.  2005, fig. 4) which show an apparent increase in $\alpha$
for $\mid l\mid >\sim 3^{\rm o}$, a change that may be removed by detailed modelling 
(Wegg \& Gerhard 2013). It should be noted that the ages of red
clump stars can cover a large range, so that the mean age of the Nishiyama
stars is not known.

In Fig.~\ref{ModLong1} for the short period stars, there is, as previously noted, no 
significant evidence of a bar. However, the structure
is complex. Apparently, the stars at lower longitudes are closer than those further out.
In discussing this (and also the long period stars) it is necessary to remember that the stars at low
longitudes are in the mean at lower latitudes (compare Fig.~\ref{allstar}). Subdivision of the Miras by both latitude
and longitude seems hardly feasible given the size of our sample. We therefore conclude that the short period
(old) group of Miras shows little evidence of barred structure but that more complex structure is evident.

The variation of bulge structure with Mira period indicates that at least
the bulk of long period Miras are not the evolutionary product of merged
binaries (and therefore belong to an old, low mass population), confirming
their intermediate age status.

\subsection{Miras at larger $\mid b \mid$} \label{largeb}
We first consider the Miras in the two SAAO fields shown in Fig.~\ref{allstar}.
In view of the discussion in the last subsection, we divided these stars
into two groups; $\log P$ greater or less than 2.6.
The results are shown in Fig.~\ref{p45Aug15} and tabulated in Table~\ref{TabRthet}.

The results are broadly similar to the inner region just
discussed. For the short period stars, the results show little evidence
of barred structure ($\alpha = 81^{\rm o} \pm 5^{\rm o}$) and the mean
moduli of the two SAAO fields differ only marginally
($\delta Mod = 0.09\pm 0.06$). For the long period group there is definite indications
of a bar ($\alpha = 62^{\rm o} \pm 14^{\rm o}$), despite the fact that the error is largely due to the relatively
small number of stars involved.

Because the SAAO fields are in outer, less dense bulge fields, the results
might be affected by foreground and background stars.  Table~\ref{trunc}
therefore lists the results obtained using a 2-$\sigma$ cut and also when
restricting the sample to Miras between distance moduli of 13 and 16. We
  have also experimented with truncating at a fixed distance either
  side of the centre. Truncating the data at $\pm 6$\,kpc about the best fit, gives results similar to
the $\pm 1.5\sigma$ truncation. Narrowing the truncation limits
reduces both $R_0$ and $\alpha$ for both the long and short period Miras,
but maintains the difference in $\alpha$ between them. The results
for the long period stars show clear evidence for a bar with $\alpha$
decreasing slightly in the restricted solutions.  For the short period stars
the restricted samples give a value of $\alpha$ significantly different from
$90{\rm ^o}$.  One interpretation of this would be a continuous variation in bulge
structure with age, the oldest Miras having little or no barred
structure, consistent with Fig.~\ref{thtlogp}.

The results for the inner bulge region tabulated in Table~\ref{TabRthet}, show a smaller value of $R_0$
for the longer period (barred system) than for the shorter period one. Some of this difference
may be due to problems of correctly  accounting for circumstellar extinction. The difference
between the short and long period stars is even more marked in the outer fields now being discussed
and it seems worth investigating whether this is connected with (3D)
bar structure.

Fig.~\ref{p45Aug15} also shows the Miras which lie at $\mid b \mid >6.0^{\rm o}$. The points at large $\mid
l \mid$ are mainly from
the SAAO fields, those near $l=0$ are from the OGLE data. The results
are shown in Table~\ref{TabRthet} and by the lines in Fig.~\ref{p45Aug15}.
They are not significantly different from those just
discussed. However, in  Fig.~\ref{p45Aug15}, the long period Miras
within $-2^{\rm o} <l< 2^{\rm o} $, fall above the mean line and have a mean distance
modulus of $14.83 \pm 0.49$  ($8.31 \pm 1.9$\,kpc). This suggests
that the low value of $R_0$ for long period stars in fields with large $l$ is at least partly due to 3D structure and perspective effects.

\subsection{Summary}

 Our results for the longer period (younger) Miras show the presence of
triaxial, bar-like, structure in both our inner and outer fields.  This
confirms the result of Whitelock \& Catchpole (1992) and Whitelock (1992)
who found a tilted bar in a sample of mainly longer period Miras.  These
were concentrated in two, one degree wide, strips at $ \mid b \mid =
7.5^{\rm o}$ extending between $-15^{\rm o} < l < 15^{\rm o} $.  The shorter
period (old) Miras in our outer fields are consistent with a spheroidal or
near-spheroidal distribution, though evidence of structure on a smaller scale
is present.  Gran et al.  (2015) suggest that their results on RR Lyrae
variables (another old population) in a field at $ l,b:-9^{\rm o},-9^{\rm
o}$ are also consistent with a near spheroidal distribution.  Furthermore,
fig. 14 of Pietrukowicz et al.  (2012) shows that the apparent distribution
of RR Lyrae variables with $|l|<3^{\rm o}$  becomes flat (larger $\alpha$) as
one moves from $ b= 2.5-4^{\rm o}$ to $ 4-5^{\rm o} $.  This same diagram
shows that, in contrast, the red clump stars show a tilted bar at all $b$ in
the plot.  This seems to suggest that the bulk of the red clump stars in
these regions belong to an intermediate age population. 
 In our inner region, the shorter period (old) Miras show a different
distance distribution from the younger ones in Fig.~\ref{ModLong1}, with no
clear evidence for a bar, though there appears to be smaller scale structure
present.

\section {Distance to the centre ($R_0$)} \label{distance}
 Glass \& Feast (1982) made a first attempt to derive a distance to the
Galactic centre from Mira variables in the NGC\,6522 and Sgr\,I bulge windows of
Baade (see also Glass et al.  1995).  The matter is complicated by the
barred nature of the bulge.  Matsunaga et al. (2009) overcame this problem by
detecting and studying Miras within 30 arcminutes of the Galactic centre. 
They obtained $8.43 \pm 0.43$\,kpc based on an LMC distance modulus of 18.50.
 Estimates derived in other ways are listed by Feast (2013). A mean value of
8.0\,kpc was adopted there with an uncertainty ``probably less than 0.5\,kpc".

The  values of $R_0$ listed in Table~\ref{TabRthet} are in reasonable
agreement with the above values, given the likely effect of 3D structure on
our values, as well as the presence of smaller scale structure
etc., discussed in the previous section. We note again that the standard errors quoted are only internal.
 
\begin{table*}
\caption{$R_0$ and $\alpha$ derived for different reddening, methods
for Miras with $ |b| < 4.5^{o}$. In the last row only Miras that lie
inside the Mira box, as defined in Fig.~\ref{jhhk}, are used. }
\label{TabRT_Red}
\begin{tabular}{lcccccc} \hline
\multicolumn{1}{c}{} & \multicolumn{3}{c}{$2.1\leq \log P<2.6$} &
\multicolumn{3}{c}{$2.6\leq \log P<2.7$} \\
 \multicolumn{1}{c}{Method} &
\multicolumn{1}{c}{$R_0$ kpc} & \multicolumn{1}{c}{\large{$\alpha$}} &
\multicolumn{1}{c}{N } &\multicolumn{1}{c}{$R_0$ kpc} &
\multicolumn{1}{c}{\large{$\alpha$}} & \multicolumn{1}{c}{N } \\
 \hline
 None        & $10.70 \pm 0.04$ & $76^o \pm 3^o$ & 4883 & $11.35 \pm
 0.12$ & $41^o \pm 4^o$ & 1182 \\
Interstellar only (Schlafly)& $9.30 \pm 0.04$ & $74^o \pm 3^o$ & 4883 & $9.87 \pm 0.11$ & $39^o \pm 4^o$ & 1182  \\ 
Whitelock  & $8.99 \pm 0.04$ & $84^o \pm 3^o$ & 4883 & $8.33 \pm
 0.08$ & $49^o \pm 5^o$ & 1182  \\
 Glass & $8.77 \pm 0.04$ & $84^o \pm 3^o$ & 4883 & $8.08 \pm 0.08$ & $ 49^o \pm 4^o$ & 1182  \\
 JHK & $8.89 \pm 0.04$ & $84^o \pm 3^o$ & 4883 & $8.04 \pm 0.08$ & $ 49^o \pm 5^o$ & 1182  \\
 \hline
\end{tabular}
\end{table*}

\begin{table*}
\caption{$R_0$ and $\alpha$ derived for different distance modulus truncations}
\label{trunc}
\begin{tabular}{lcccccc} \hline
\multicolumn{1}{c}{} & \multicolumn{3}{c}{$2.1\leq \log P<2.6$} &
\multicolumn{3}{c}{$2.6\leq \log P<2.7$} \\
 \multicolumn{1}{c}{Method} &
\multicolumn{1}{c}{$R_0$ kpc} & \multicolumn{1}{c}{\large{$\alpha$}} &
\multicolumn{1}{c}{N } &\multicolumn{1}{c}{$R_0$ kpc} &
\multicolumn{1}{c}{\large{$\alpha$}} & \multicolumn{1}{c}{N } \\
 \hline
\multicolumn{7}{c}{ $-4.5 < Latitude < +4.5$}\\
 $13 < (m-M)_0 < 16 $     & $8.73 \pm 0.03$ & $79^o \pm 3^o$ & 4671 & $8.18 \pm
 0.07$ & $56^o \pm 5^o$ & 1132 \\
 $ 1.5 \sigma$  & $8.67 \pm 0.02$ & $76^o \pm 2^o$ & 4326 & $8.07 \pm
 0.08$ & $47^o \pm 3^o$ & 1038  \\
 $ 2.0 \sigma$ & $8.77 \pm 0.03$ & $77^o \pm 3^o$ & 4607 & $8.17 \pm 0.06$ & $
 49^o \pm 4^o$ & 1107  \\
No limit & $8.99 \pm 0.04$ & $84^o \pm 3^o$ & 4883 & $8.33 \pm 0.08$ & $
 49^o \pm 4^o$ & 1182  \\
\multicolumn{7}{c}{ SAAO Fields Only}\\
$13 < (m-M)_0 < 16 $  & $8.40 \pm 0.10$ & $75^o \pm 4^o$ & 486 & $7.62 \pm 0.24$ & $55^o \pm 8^o$ & 85  \\ 
$ 1.5 \sigma$ & $8.50 \pm 0.09$ & $70^o \pm 3^o$ & 464 & $7.55 \pm 0.23$ & $ 50^o \pm 6^o$ & 84  \\
$ 2.0 \sigma$ & $8.50 \pm 0.11$ & $73^o \pm 4^o$ & 491 & $7.55 \pm
0.25$ & $ 53^o \pm 8^o$ & 88  \\
No limit & $8.26 \pm 0.14$ & $82^o \pm 6^o$ & 525 & $7.14 \pm 0.94$ & $
 62^o \pm 14^o$ & 94  \\ 
 \hline
\end{tabular}
\end{table*}

 Table~\ref{TabRT_Red} shows how the derived values of $R_0$ and $\alpha$ vary using different
 methods of correcting for interstellar and circumstellar
 extinction. 
 `None' indicates that no extinction correction was made.
The main significance of this table is to show that values of $\alpha$ are insensitive to the 
extinction corrections used. As expected, neglecting circumstellar extinction leads to
an overestimate of $R_0$, especially for the longer period stars.
 Using the ``Matsunaga" method (section \ref{ext}) leads to $R_0 = 8.9$\,kpc for the 
short period
group in Table~\ref{TabRT_Red} and $ 8.5$\,kpc for the long period group, in good agreement
with the ``Whitelock" results in that table. The difference between these two values is probably a consequence of uncertainties in the reddening correction (see section 6) and/or due to the 3-dimensional structure (see section 8) of the bulge.
 Since the ``Whitelock" results are in the
SAAO system and the ``Matsunaga" results in the 2MASS system, the agreement suggests that
conversion between the systems has little effect on the results.

It is evident from, for example Fig.~\ref{ModLong},  that there is a
considerable apparent spread in distance moduli of the Miras in any one
direction.  Part of this is no doubt real and part due to uncertainty in
the derived moduli.  The effects of truncating the sample in the SAAO fields
was already discussed in section \ref{largeb}.  The first part of Table 4
shows the effects of similar experiments with the lower latitude stars.  The
effects are relatively small, especially considering that the
uncertainties in the results are dominated by external errors (see
following paragraph). Truncation has the effect of slightly reducing the value of $R_0$ for the Miras at low latitude and increasing them for the high latitude SAAO fields,
presumably because of the different balance between foreground and
background objects.

Matsunaga et al. (2009) give some discussion of the real (external) errors
of their value of $R_0$ derived from Miras in the region of the
centre. They adopt 0.4\,kpc for this quantity. Our Miras suffer less
interstellar absorption than theirs but are further from the centre
and are thus not fully model independent. This is particularly true of
the longer period stars which are in a bar structure. The correction
for circumstellar extinction is also more important for these latter
stars. We thus adopt as our best estimate of $R_0$ the value for Miras
with $\log P < 2.6$, restricted to $|b|<4.5^{\rm o}$
and taking a mean of the last three estimates in Table~\ref{TabRT_Red}
(8.9\,kpc) and adopt Matsunaga's error estimate (0.4\,kpc).

\section{ Structures within the Bar} 

 Whether on not there is small scale structure in the bulge is of
considerable significance for its formation and evolution.

To illustrate possible structure, we have selected Miras within a restricted range of $l$ and $b$ over which
the coverage is fairly uniform, as can be judged by looking at
Fig.~\ref{allstar}. The Miras are projected onto the Galactic,
$b=0^{\rm o}$ plane, after
correcting for ``Whitelock" extinction. The plane is divided into 0.15\,kpc square bins and
the number of Miras is counted in each bin. Each bin becomes an image
pixel
and the maximum number of stars in a pixel is 15 in the short period
group and 7 in the long.

\begin{figure} 
\includegraphics[width=8.5cm,angle=0]{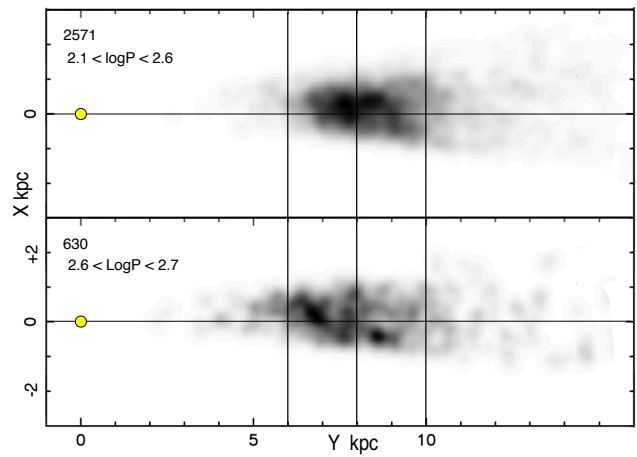}
 \caption{Distribution of the long and short period Miras, projected on
to the Galactic plane showing clumpiness, with the Sun marked at 0,0. Miras are counted within 0.15\,kpc square
   bins, then smoothed and each plot normalised to the bin with the maximum number
   of stars. The plots are for
   $-6^{\rm o}<l<10^{\rm o}$ and $-4.5^{\rm o}<b<-2.5^{\rm o}$, over
   which the areal coverage is nearly uniform. The numbers are the
   total number of Miras in the image.} 
 \label{miraclump}
\end{figure}

 The maximum number of
stars is used to scale the resulting image and normalise it
to an intensity count of 255. The image is then smoothed and doubled
in size
to make it more legible. Images for the short and long period Miras
are shown in Fig.~\ref{miraclump}. The restriction in $l$ and $b$ means that the images do not correspond to
the subsets of data given in Table~\ref{TabRthet}.

In Fig.~\ref{miraclump} there is a strong contrast between the
spatial distribution of old $2.1\leq \log P<2.4$ and intermediate age  
$2.6\leq \log P<2.7$ Miras.
The short period  group show a relatively smooth centrally
concentrated distribution. The long period group show a clumpy structure
as well as evidence of the tilted bar to which the two main clumps at 6.8
and 8.6\,kpc seem to belong.

\begin{figure} 
\includegraphics[width=8.5cm,angle=0]{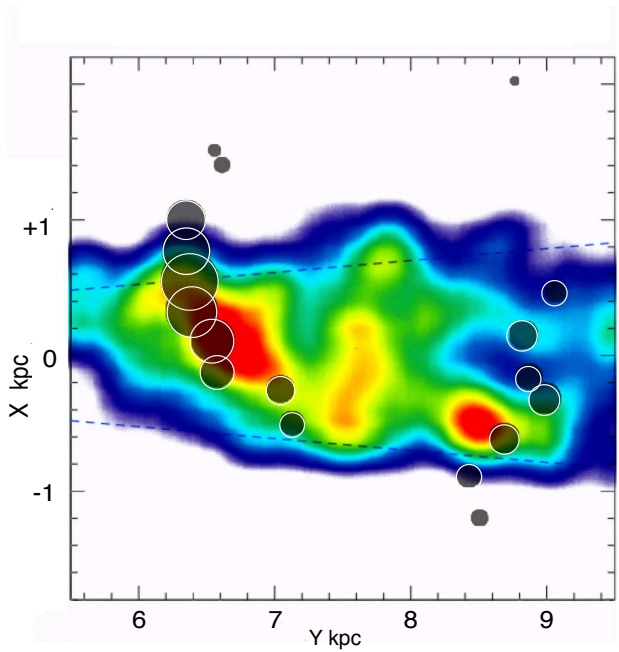}
 \caption{The distribution of Miras (false colour with red the maximum) with,$-6^{\rm o}<l<10^{\rm o}$ and
   $-4.5^{\rm o}<b<-2.5^{\rm o}$ and $2.6\leq \log P<2.7$, overlain on
   fig. 6 (grey circles) of McWilliam \& Zoccali (2010), showing
   peaks in their red clump stars. Their distance scale is
   matched to ours.}
 \label{mirasmoothMcW}
\end{figure}

In Fig.~\ref{mirasmoothMcW} the clumping in the long period Miras is
compared with the distribution of red clump stars found by McWilliam \&
Zoccali (2010) in the $b=-8^{\rm o}$ plane, shown in their fig.  6.  They
interpret these concentrations as the wings of an X shaped or triaxial bar,
seen almost edge on to our line of sight, though this
interpretation has been disputed (Lee, Joo \& Chung 2015).
The correspondence between some of their clumps and ours, shown in Fig.~\ref{mirasmoothMcW}, suggests that
the red clump stars and the long period Miras have similar distributions.
Taken together with the results of the previous paragraph, this would
 indicate a more clumpy structure in a (younger) triaxial or X-shaped bar
than in the older part of the bulge. It has been proposed
that the bulge formed from the infall and eventual merger of large massive lumps
(e.g. Noguchi 1999, Immeli et al. 2004, Ferraro et al. 2009, Massari et al. 2015)
whether they could be long enough lasting  to explain the observed clumping
is not clear.

When looking at the structures, shown in Fig.~\ref{miraclump}
   and Fig.~\ref{mirasmoothMcW}, it is important to remember that they lie within
 a slice, which passes through the centre of the Galaxy between 350pc
and 600pc above the Galactic plane.

\section{Conclusion }

We have obtained periods and $JHKL$ photometry for 648 Mira variables in two
outer fields in the direction of the Galactic bulge and combined it with
8057 Miras found in the OGLE and Macho surveys of the inner bulge. 
Distances were derived from a period-luminosity relation together with
corrections for interstellar and circumstellar extinction.  In both our
inner and outer regions the longer period Miras, tracers of an intermediate
age population, give evidence of a tilted bar-like structure. The
older, shorter period Miras do not show this. The longer period Miras
also show a more clumpy structure than the shorter period ones, possibly part
of an X-structure.   

We estimate a distance to the Galactic centre in satisfactory
agreement with other estimates.

The change of Mira distribution with period, besides showing the complex
structure of the bulge, also indicates that the longer
period stars there, are fundamentally different from the shorter period ones.
It therefore supports the view that the longer period Miras in the bulge
belong to an intermediate age population, as such stars do elsewhere, and are
not, for instance, the progeny of the mergers of old (low mass) stars. The exact age
of that intermediate age population may depend on abundance details and particularly on the level 
of helium enrichment.

\section{Acknowledgements}  

We are grateful to the following people who took part in the infrared
photometry programme at Sutherland, Brian Carter, Dave Laney and
Enrico Olivier. MWF and PAW gratefully acknowledge a research grant
from the SA National Research Foundation. We thank an anonymous 
referee for their comments and specifically for raising the question of the helium abundance.
This paper uses observations made at the South African Astronomical Observatory (SAAO).


\end{document}